\documentclass[reprint,superscriptaddress,amsmath,amssymb,prl,floatfix]{revtex4-2}
\usepackage{graphicx}
\usepackage{floatrow}
\usepackage{subfig}
\usepackage{dcolumn}
\usepackage{bm}
\usepackage[normalem]{ulem} 
\usepackage[makeroom]{cancel}
\usepackage{xcolor}
\usepackage{hyperref}
\hypersetup{colorlinks=true, citecolor=blue, urlcolor=blue, linkcolor=blue}

\begin{document}
\preprint{XXXX}

\title{How Social Rewiring Preferences Bridge Polarized Communities}
\author{Henrique M. Borges}
    \email{henrique.joao.machado.borges@tecnico.ulisboa.pt}
    \affiliation{Departamento de Física, Instituto Superior Técnico, Universidade de Lisboa, Lisbon, Portugal}
\author{Vítor V. Vasconcelos}%
    \email{v.v.vasconcelos@uva.nl}
    \affiliation{Computational Science Lab, Informatics Institute, University of Amsterdam, Amsterdam, The Netherlands}
    \affiliation{POLDER, Institute for Advanced Study, University of Amsterdam, Amsterdam, The Netherlands}
    \affiliation{Center for Urban Mental Health, University of Amsterdam, Amsterdam, The Netherlands}
\author{Flávio L. Pinheiro}%
    \email{fpinheiro@novaims.unl.pt}
    \affiliation{NOVA Information Management School (NOVA IMS), Universidade Nova de Lisboa, Lisbon, Portugal}
\date{\today}

\begin{abstract}
 Recently, social debates have been marked by increased polarization of social groups. Such polarization not only implies that groups cannot reach a consensus on fundamental questions but also materializes in more modular social spaces/networks that further amplify the risks of polarization in less polarizing topics. How can network adaptation bridge different communities when individuals reveal homophilic or heterophilic social rewiring preferences? Here, we consider information diffusion processes that capture a continuum from simple to complex contagion processes. We use a computational model to understand how fast and to what extent individual rewiring preferences bridge initially weakly connected communities and how likely it is for them to reach a consensus. We show that homophilic and heterophilic rewiring have different impacts depending on the type of opinion spread. First, in the case of complex opinion diffusion, we show that even polarized social networks can reach a population-wide consensus without reshaping their underlying network. When polarized social structures amplify opinion polarization, heterophilic rewiring preferences play a key role in creating bridges between communities and facilitating a population-wide consensus. Secondly, in the case of simple opinion diffusion, homophilic rewiring preferences are more capable of fostering consensus and avoiding a co-existence (dynamical polarization) of opinions. Hence, across a broad profile of simple and complex opinion diffusion processes, only a mix of heterophilic and homophilic rewiring preferences avoids polarization and promotes consensus.
\end{abstract}
\maketitle

\section{Introduction}
In the past decades, social media platforms have been at the center stage of social debate and occupy a major role in our social dynamics. These platforms have also amplified our tendency to form polarized groups whose segregation and clustering of views prevents them from reaching consensus even on the most fundamental societal questions \cite{vasconcelos2021segregation}. As such, it is not surprising that much research has been conducted to understand the phenomena of social polarization better \cite{garimella2018quantifying, chitra2020analyzing, jacob2023polarization,kubin2021role, barbera2020social, sunstein2018social}. While past works focused on identifying underlying mechanisms that can lead to social polarization, both in opinion composition \cite{kozma2008consensus, nardini2008s, krueger2017conformity, sirbu2019algorithmic} and in respect to the structural organization of communities \cite{durrett2012graph, sasahara2021social, yu2017opinion, peralta2021effect, santos2021link,o2015mathematical}, few works have looked into how dynamical processes on already structurally polarized populations can amplify or mitigate the degree of structural polarization of a community. Here, we study how the co-evolution of an information diffusion process---that interpolates between simple and complex contagion processes---and the network structure of an initially polarized social network can lead to the reshaping of social structures and build environments that are more suitable for the formation of consensus.

Empirical evidence suggests that distinct types of information spread differently 
\cite{goffman1964generalization, daley1964epidemics,centola2010spread, sprague2017evidence, monsted2017evidence, vespignani2012modelling, aral2009distinguishing}, but that there is a positive and direct relationship between the probability that an individual adopts new information and the number of friends that already hold it \cite{karsai2014complex, bakshy2012role, backstrom2006group, bakshy2009social, cha2009measurement, sprague2017evidence, monsted2017evidence}. In that context, information diffusion models can be divided into simple contagion (social learning) or complex contagion (social influence) processes. Formally, in simple contagion processes, the probability that information is transmitted is directly proportional to the fraction of neighbors with such information \cite{goffman1964generalization, kempe2003maximizing, daley1964epidemics,centola2010spread,valente1996social}. In contrast, under complex contagion, adoption is typically modeled using a threshold function, where the probability of transmission is one if the fraction of neighbors with that information exceeds a given threshold and zero otherwise \cite{granovetter1978threshold, centola2018behavior, monsted2017evidence, lehmann2018complex}. However, empirical evidence supports the view that a heterogeneous distribution of thresholds better describes populations \cite{karsai2016local, karsai2014complex}, leading to the proposal of more general models \cite{vasconcelos2019consensus,horsevad2022transition}.

\begin{figure*}[!t]
      \includegraphics[trim={0cm 4.75cm 0 0},width=1.\textwidth]{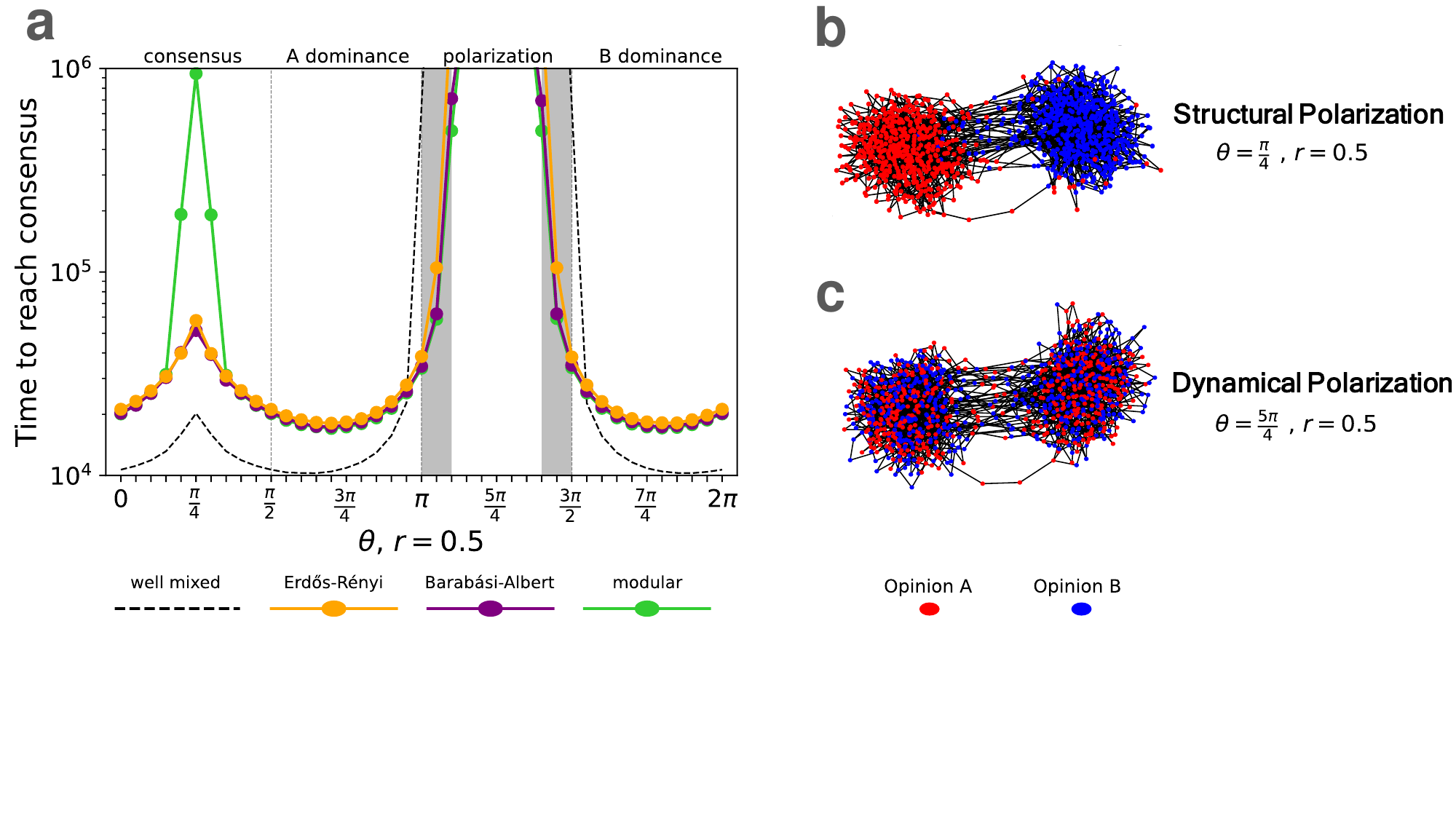}
    \caption{\textbf{Opinion dynamics in static social networks}. Panel \textbf{a} shows the fixation times, measured for fully connected communities (well-mixed, black dashed lines) and structured populations (different topologies in orange, purple, and green lines). These results averaged over $1.0 \times10^4$ independent simulations starting from a configuration with equal abundances of opinions. Vertical lines separate the different dynamical regions described in the main text, and gray areas indicate the mismatch between well-mixed and structured populations. This figure corresponds to $p=0$, matches Figure 3a from Ref. \cite{vasconcelos2019consensus}, and sets up a baseline scenario in the absence of rewiring. Panels \textbf{b} and \textbf{c} illustrate the possible outcomes of structural and dynamical polarization. The former is characterized by a scenario in which polarization occurs due to structural lock-ins. In the latter, polarization results from agents' inability to reach a consensus due to co-existence-like dynamics.}
    \label{fig:Static_Scenario_Figure}
\end{figure*}

While, from an information diffusion perspective, polarization can be characterized by a population that cannot reach a consensus (i.e., the majority of the population cannot align towards the same opinion), structurally speaking, a polarized population can be described by a modular network structure with dense within- and sparse between-community connections. Modular structures emphasize the amplification of group differences in terms of complex information (i.e., complex contagion or social influence) but do not affect the dissemination of simple information (i.e., simple contagion or social learning) \cite{centola2007complex,weng2013virality}. To break such structural lock-ins, populations must reshape their connections. In that sense, two adaptive network mechanisms stemming from individual choices are homophily---the degree to which individuals desire similarity between social contacts---and heterophily---desiring difference. Coupling the agents' dynamic states and connections leads to a feedback loop where the network structure and individuals' opinions affect each other. Past works proposed models that combine opinion dynamics with homophilic and heterophilic network dynamics \cite{holme2006nonequilibrium, kimura2008coevolutionary, Vazquez2008}, but they have not addressed the interplay between the type of information diffusion and the rewiring mechanism taking place. 

Here, we study the feedback between the dynamics of individual opinions and network structure in contemporary (polarized) social networks and ask to what extent heterophilic and homophilic individual rewiring preferences can lead to the bridging of initially polarized communities. We focus on potential future debates, which will exhibit a range of diffusion properties, and test how different rewiring preferences influence the potential for consensus formation in competitive opinion dynamics. Furthermore, we show how the resulting network structures put populations at risk of polarization in future social debates. Across a broad profile of simple and complex opinion diffusion processes, only a mix of heterophilic and homophilic rewiring preferences avoids polarization and promotes consensus.


\section{Materials and Methods}
\label{section:methods}
Let's consider a finite but large population of $Z>>1$ individuals where each agent is characterized by one of two contrasting opinions, $A$ or $B$. At any given moment, the population contains a fraction $x = n^A/Z$ of \emph{A}s and $1-x = n^B/Z$ of \emph{B}s. Moreover, individuals are embedded in a complex network of social relationships, where each node corresponds to an individual, and links capture who influences whom. 

We study the case of a co-evolving population in which individuals can update their opinions and adapt their ties. We consider a stochastic one-step process in which, at each time step, one of two events takes place. With probability $p$, individuals attempt to rewire a social tie, and, with probability $1-p$, their opinion. In both cases, the decision depends on the composition of individuals' neighborhoods.

\begin{figure*}[!t]
    \includegraphics[trim={1cm .25cm 4.65cm 0},width=0.8\textwidth]{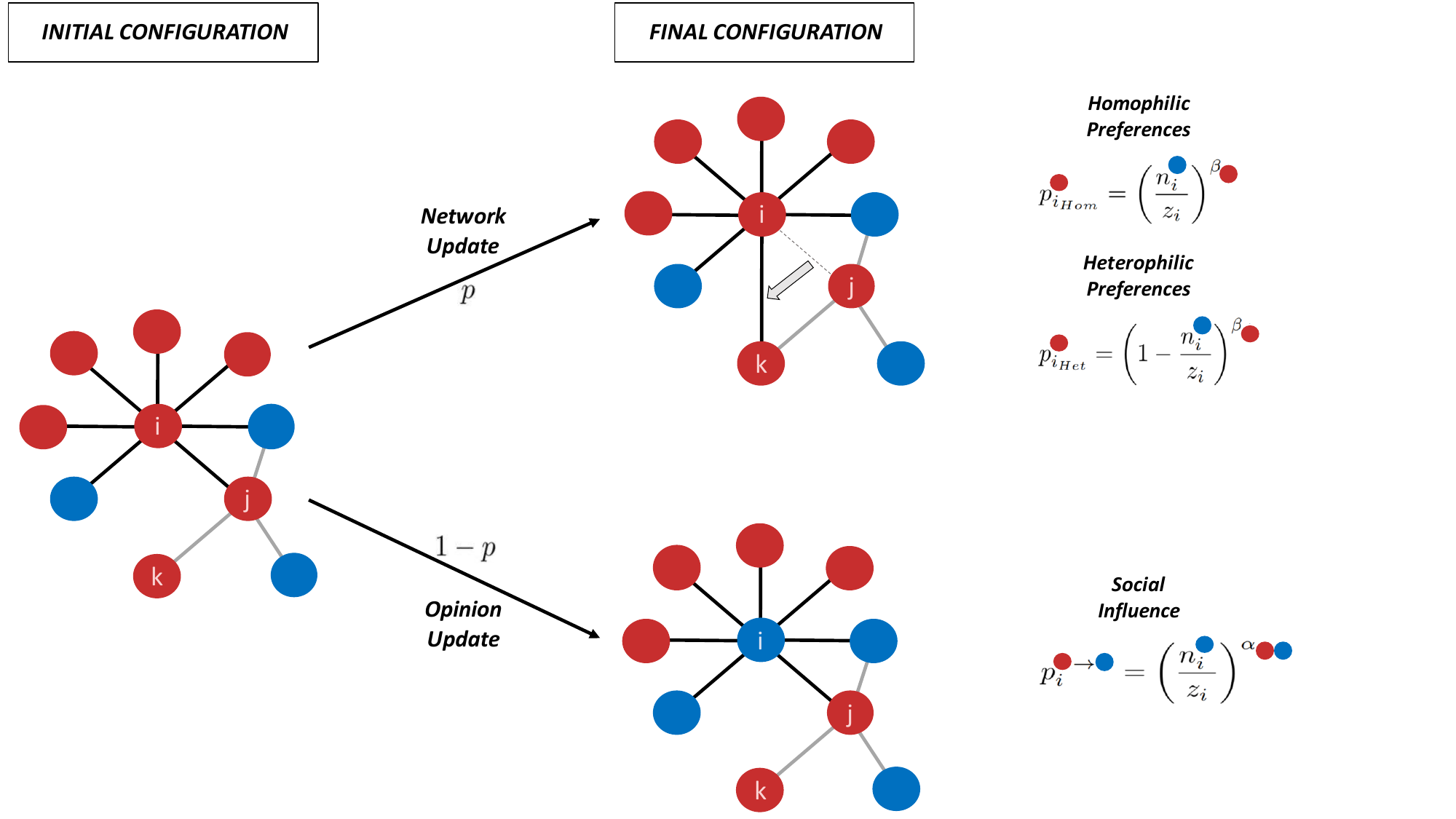}
    \caption{\textbf{Schematic illustration of the co-evolutionary model used in this manuscript}. The proposed model combines a competitive opinion diffusion process that co-evolves with a network dynamics process that can follow homophily (individuals have a preference to be connected with individuals of the same opinion) or heterophily (individuals have a preference to be connected with individuals of opposite opinion).}
    \label{fig:Schematic_Figure_Model}
\end{figure*}

\subsection{Opinion Dynamics}
During an opinion update step, an agent $i$ is selected at random and updates its opinion $X\in\{ A, B\}$ to $Y\in\{ A, B\}$ according to 
\begin{equation}
    p_i^{X \to Y}=\bigg(\frac{n^Y_i}{z_i}\bigg)^{\alpha_{XY}},
\end{equation}
where $z_i$ is the degree of the individual $i$, $n^Y_i$ is the number of $i$'s neighbors with opinion $Y$, and $\alpha_{XY} \geq 0$ is the complexity of opinion $Y$ when learned by an individual with opinion $X$. When $\alpha_{AB} = \alpha_{BA} = 1$, the model resumes to the voter's model. It is convenient to reparametrize the complexities into polar coordinates, such that $\alpha_{AB}=1+r\sin \theta$ and $\alpha_{BA}=1+r\cos \theta$. Hence, with a single parameter $\theta$, we can explore the four dynamical regions of interest in fully connected populations:
\begin{itemize}
    \item   \textbf{A dominance}, for $ \pi/2 \leq \theta < \pi $:  This region does not have any internal fixed point and $x^*=0$ is unstable and $x^*=1$ is stable. Opinion A will dominate the population. 
    \item  \textbf{B dominance}, for $ 3\pi/2 \leq \theta < 2\pi $:  This region does not have any internal fixed point and $x^*=0$ is stable and $x^*=1$ is unstable. Opinion B will dominate the population. In Evolutionary Game Theory (EGT), this and A dominance exhibit dynamics akin to the Prisoner's Dilemma and Harmony Game \cite{rapoport1965prisoner,licht1999games, cressman2003evolutionary}.
    \item  \textbf{Polarization}, for $\pi/2  \leq \theta < 3\pi/2 $: this region is characterized by a single stable internal fixed point that leads to the polarization of opinions, which is identified by the constant co-existence of both opinions and the inability of the population to reach a population-wide consensus due to a dynamical lock. In EGT, this outcome is dynamically similar to a 2-person Snowdrift Game \cite{doebeli2005models}.
    \item \textbf{Consensus}, for $0 \leq \theta < \pi/2$: this region has a single unstable internal fixed point resulting in coordination dynamics and a population-wide consensus, which only depends on the initial abundance of opinions. In EGT, this outcome is dynamically similar to a 2-person Stag-Hunt Game \cite{skyrms2004stag}.
\end{itemize}

In the Polarization and Consensus regions, the internal fixed point position is independent of $r$ and depends only on the ratios of complexities \cite{vasconcelos2019consensus} according to $x^\ast= \cot \theta =(\alpha_{BA}-1)/(\alpha_{AB}-1)$. For the remainder of the manuscript, we shall consider the space spanned by $r=1/2$ and $0 \leq \theta \leq 2\pi$. Figure \ref{fig:Static_Scenario_Figure} compares the fixation times (time to consensus) obtained for three different network structures across the four regions of interest. Except for modular networks, qualitatively, the expected time to reach consensus in structured populations is consistent with the well-mixed scenario. In structured populations, the Polarization region is reduced. Moreover, fixation times peak in the Consensus region in modular population structures. Such scenarios result from each community reaching a different local consensus and then being unable to converge to a population-wide consensus due to imposed structural lock-ins, a well-known result in the context of complex contagion \cite{centola2007complex}.

\subsection{Rewiring Preferences}
We consider two different families of rewires based on how individuals' networks are assessed: Homophilic or Heterophilic updates, in which individuals rewire a connection if their neighborhood is too dissimilar or similar to them, respectively. As such, during a link update step, a random individual $i$ breaks a random tie with a probability given by:
\begin{subequations}
    \begin{align}
            & p_{i_{Hom}}^X =\bigg(\frac{n_i^Y}{z_i}\bigg)^{\beta_{X}} \text{or} \label{homoph} \\ 
            & p_{i_{Het}}^X =\bigg(1-\frac{n_i^Y}{z_i}\bigg)^{\beta_{X}} \label{hetero},
    \end{align}
\end{subequations}
where $\beta_X$ accounts for the tolerance of an agent with opinion $X\in\{A,B\}$ regarding the composition of its neighborhood for the homophilic \ref{homoph} and heterophilic \ref{hetero} cases. If a link is broken, then $i$ creates a new tie with a random friend of a neighbor. This ensures that the network remains connected. 
This evaluation procedure is similar in spirit to the Schelling model \cite{schelling2006micromotives}, especially taking into account a heterogeneous-thresholds interpretation \cite{vasconcelos2019consensus}.

\begin{figure*}[!t]
    \includegraphics[width=\textwidth]{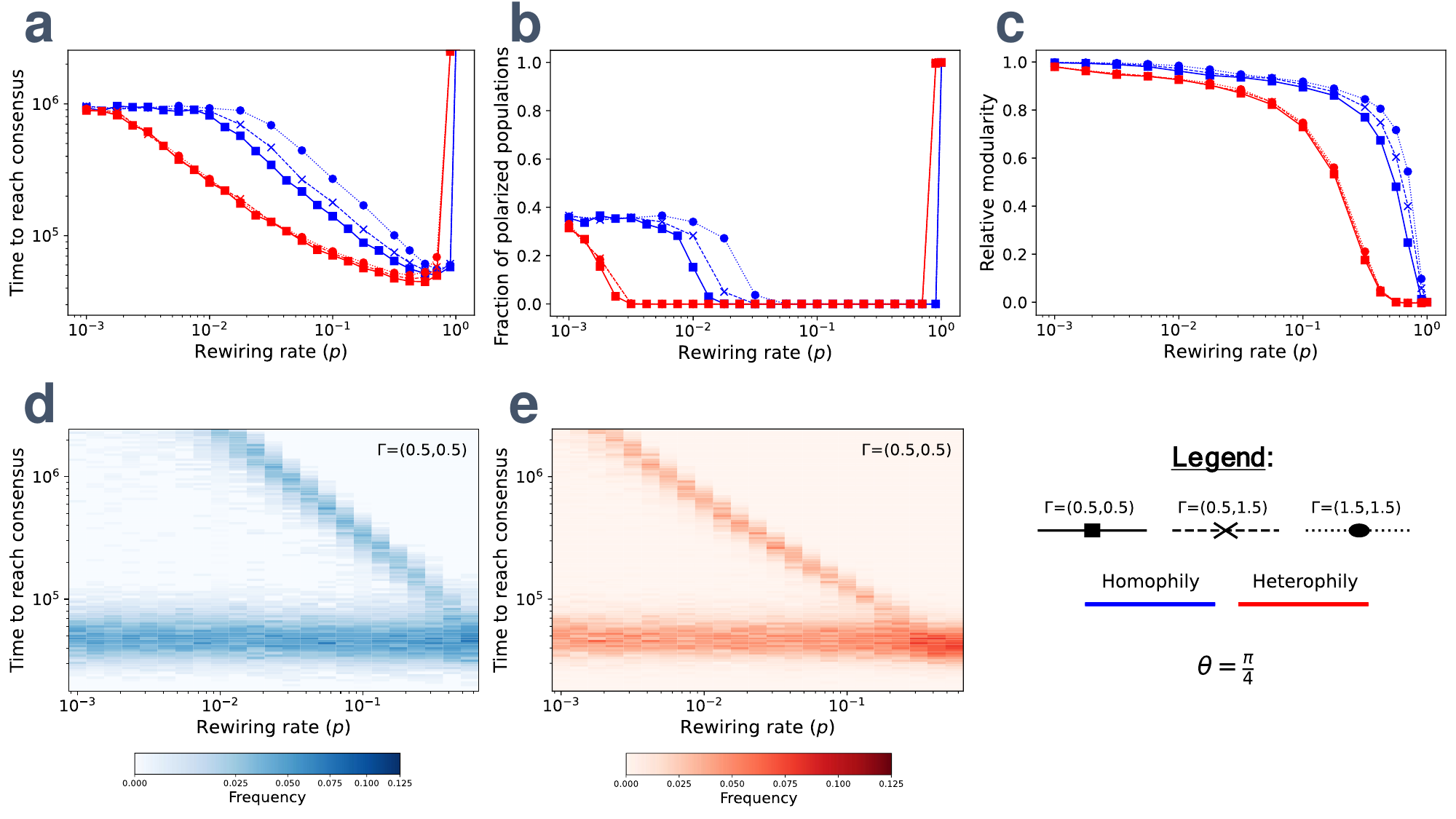} 
    \caption{\textbf{Opinion dynamics with Homophilic and Heterophilic rewiring preferences in a Consensus regime ($\theta=\pi/4$)}.  Panel \textbf{a}) shows the fixation times as a function of the rewiring rate, $p$. Panel \textbf{b}) shows the fraction of times a population ends in an opinion polarization state as a function of the rewiring rate, $p$. Panel \textbf{c}) shows how the modularity of the initial structure decays as a function of $p$. Panels \textbf{d})  and \textbf{e}), show the distribution of the fixation times until reaching consensus for different rewiring rates, $p$, for a homophilic (d) and heterophilic (e) rewiring preferences and $\beta=(0.5,0.5)$. In panels \textbf{a}), \textbf{b}), and \textbf{c}) symbols indicate different values of the strictness of rewiring decisions ($\beta$) and colors the different rewiring preferences (homophilic in blue and heterophilic populations in red). Results are the average over $10^3$ independent simulations, each with an upper bound of $2.5 \times10^6$ iterations, for an initial random distribution of equal proportions of \emph{A}s and \emph{B}s, on modular networks with $N=1000$ nodes, and average degree $\langle z_i \rangle=4 $.}
    \label{fig:Random_structural_polarization}
\end{figure*}

The tolerance coefficient, $\beta_X$, follows a similar definition of $\alpha_{XY}$. As such, it allows us to interpolate between distinct scenarios. Lower values of $\beta_X$ are associated with harsher evaluations, which makes it more likely for their neighborhood to change, compared to an individual with an identical neighborhood but a less stringent assessment (larger values of $\beta_X$). Moreover, individuals of different opinions can have different tolerance levels, $\beta_X$. As such, we define the pair of tolerance-to-rewiring coefficients as $\beta=(\beta_A,\beta_B)$.

\subsection{Simulations}
We consider the case of a population whose initial structure is modular and defined by two weakly connected communities. We generated these networks by randomly linking $q = 20$ nodes from two independently generated Barabási-Albert \cite{barabasi1999emergence} networks with $N/2$ nodes each. We considered $N=10^3$ and an average degree of $\langle z_i \rangle=4$ unless specified otherwise.

Each simulation starts with an equal proportion of \emph{A}s and \emph{B}s.  It is also assumed that all individuals have either a homophilic or heterophilic evaluation of their neighborhood. While consensus is eventually reached in finite populations, the time taken can be exceedingly long. For that reason, we set an upper bound of $M_\text{iter}=2.5 \times10^6$ iterations, which we take as the maximum time to consensus. We present the average out of $10^3$ independent simulations for each parameter set. To capture different future polarizing topics, we consider a scenario in which opinions are randomly distributed in the population and another where opinions are associated with specific modules of the network.  


\section{Consensus Regime}
Let us start by considering the properties of the diffusion process that lie in the structural polarization region for static networks when the modularity of the networks breaks their ability to reach a population-wide consensus, i.e., $\theta=\pi/4$. Figure \ref{fig:Random_structural_polarization}\textbf{a} shows the average fixation time as a function of the rewiring rate $p$, and Figure \ref{fig:Random_structural_polarization}\textbf{b} shows the fraction of simulations in which populations end polarized in terms of opinions. Besides considering populations with heterophilic (red) and homophilic (blue) rewiring preferences, we also look into the strictness of the tolerance-to-rewire coefficient (symbols), $\beta=(\beta_{A}, \beta_{B})$.

\begin{figure*}[!t]
    \includegraphics[width=\textwidth]{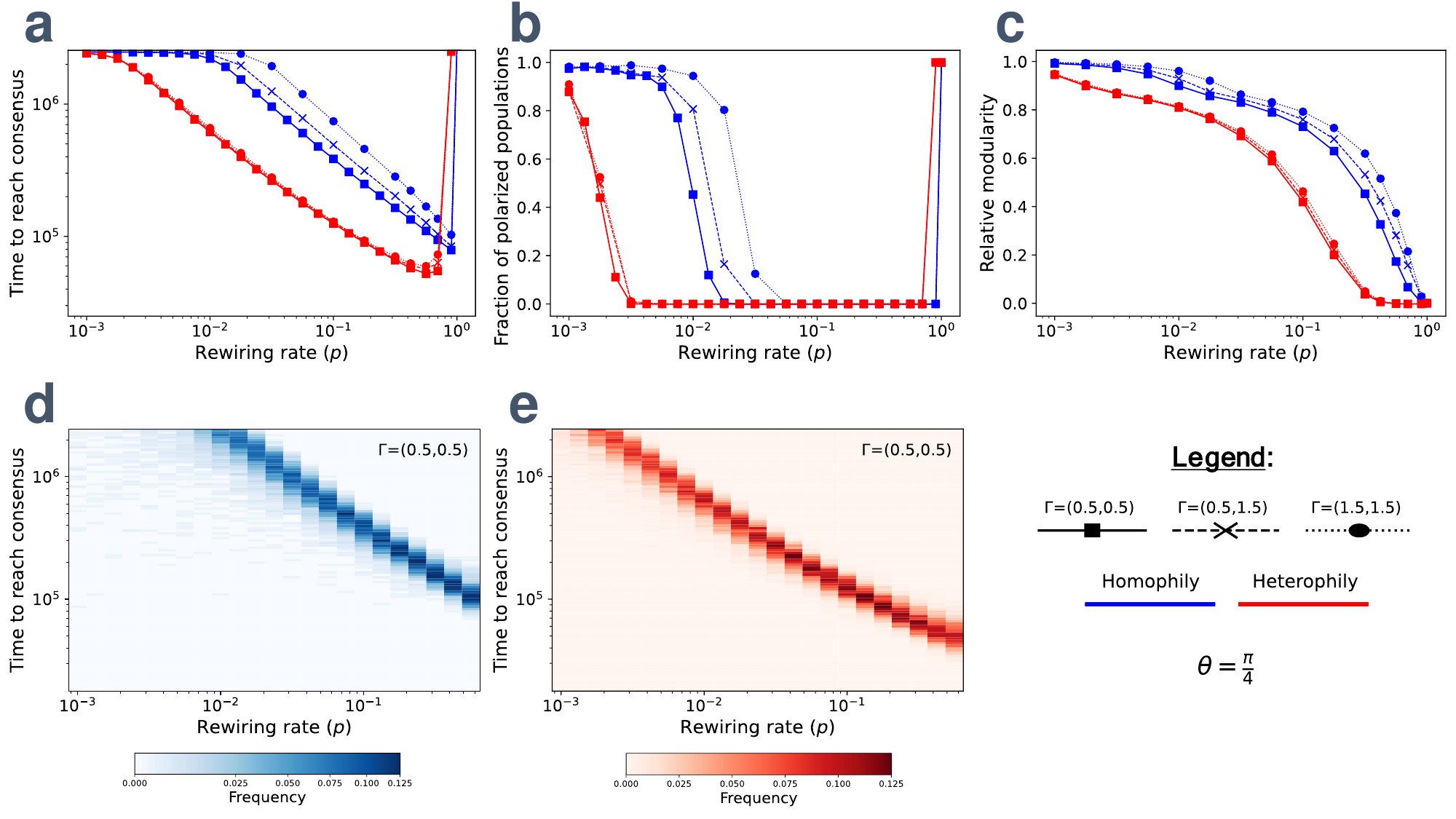}
    \caption{\textbf{Opinion Opinion dynamics with Homophilic and Heterophilic rewiring preferences in a Consensus regime ($\theta=\pi/4$) with initial opinion polarized configurations}. In this series of results, each network community starts with a local consensus on a different opinion. Panel \textbf{a} shows the fixation time as a function of the rewiring rate, $p$. Panel \textbf{b}  shows the fraction of times a population ends in an opinion polarization state as a function of the rewiring rate, $p$. Panel \textbf{c} shows how the modularity of the initial structure decays as a function of $p$. Panels \textbf{d})  and \textbf{e}), show the distribution of the fixation times until reaching consensus for different rewiring rates, $p$, for a homophilic (d) and heterophilic (e) rewiring preferences and $\beta=(0.5,0.5)$. In panels \textbf{a}), \textbf{b}), and \textbf{c}) symbols indicate different values of the strictness of rewiring decisions ($\beta$) and colors the different rewiring preferences (homophilic in blue and heterophilic populations in red). Results are the average over $10^3$ independent simulations, each with $M_\text{iter}=2.5 \times10^6$ iterations, on modular networks with $N=10^3$ nodes, an average degree of $\langle z_i \rangle=4 $.}
    \label{fig:Ordered_structural_polarization}
\end{figure*}

For low rewiring rates ($p < 10^{-3}$), we observe a flat fixation time (see Figure \ref{fig:Random_structural_polarization}\textbf{a}). In that range, the rewiring dynamics are not impactful enough to generate timely structural changes in the population structure. As such, population outcomes can be divided into two cases: the fraction of simulations in which population-wide consensus is reached in a relatively short time ($\approx 10^{5}$ iterations) since both communities reach the same consensus independently and those simulations in which each community reaches a different local consensus and for which the simulations stop at the designated $M_\text{iter}$ iterations. Hence, the initial plateau observed is not the maximum number of iterations but an average of the combination of times from those two scenarios. 

In the consensus regime, for intermediate values of rewiring rates ($10^{-3} < p < 10^{-1}$), the nature of the assessment (homophily vs heterophily, color) has a more significant impact on the results than the strictness of the assessment (tolerance level, symbols). Homophilic rewiring preferences show decreased fixation times with lower tolerance-to-rewrite coefficients ($\beta$), whereas heterophilic rewiring leads to shorter fixation times overall. 

The rewiring rate determines if structurally polarized populations can reshuffle their social structure and bridge communities in due time and, thus, foster a population-wide consensus (see Figure \ref{fig:Random_structural_polarization}\textbf{b}). As such, the average time required to reach a consensus decreases monotonically with $p$. However, as shown in Figures \ref{fig:Random_structural_polarization}\textbf{d} and \ref{fig:Random_structural_polarization}\textbf{e}, the average time will continue to have two distinct contributions: one that represents the scenario where both modules reach consensus ($\approx 10^{-5}$) and another corresponding to situations where consensus is achieved through the rewiring of links and the break of the initial polarized social structure. 

It is possible to track the degree by which rewiring reshapes the original network by tracking how the network modularity \cite{newman2003mixing,newman2004finding,newman2006modularity} decays with $p$ (see Figure \ref{fig:Random_structural_polarization}\textbf{c}). We compute the modularity assuming the two initial communities as the network partitions. For low values of rewiring rate ($p$), the final networks can still keep their initial modular structure intact. However, when the rewiring probability is large, the network structure begins to lose its distinctive modular properties, initially slowly and then abruptly. In fact, it is possible to observe a sharp transition in the relative modularity of the network at a critical rewiring probability (see Figure \ref{fig:Random_structural_polarization}\textbf{c}). This critical point marks a transition between a `modular-like network phase,' below the critical rewiring rate $p_c$, and a `randomly rewired network phase,' for $p > p_c$, where the final networks lose the initial modular character into that of a completely mixed structure with the number of edges between each pair of nodes (within one of the original communities) being equivalent to that of a network that has undergone random rewiring and shares the same degree distribution with this final network. 

Overall, the dynamics can be separated into two distinct phases: an initial fast convergence to population-wide consensus and a second case that lasts longer and in which the population is first stuck in a polarized opinion state along the community structure of the network and then, through link rewiring, is able to build the necessary bridges to reach consensus. For that reason, we investigate what occurs when the population starts from an initial condition of an opinion-polarized population with local consensus along the network's community structure. Since each community makes up half of the population, we start with the same abundance of opinions as before. Figures \ref{fig:Ordered_structural_polarization}\textbf{a} and \ref{fig:Ordered_structural_polarization}\textbf{b} show that for  $p<p_c$, fixation times are longer for this initial set-up in relation to a random initial set-up, and, for $p \gtrsim 0$, the average fixation time is $M_\text{iter}$ and, thus, all populations end in a polarized state, effectively removing the scenario where the network achieves fast population-wide consensus (see, Figures \ref{fig:Ordered_structural_polarization}\textbf{d} and \ref{fig:Ordered_structural_polarization}\textbf{e}). Moreover, while Figures \ref{fig:Ordered_structural_polarization}\textbf{a} and \ref{fig:Ordered_structural_polarization}\textbf{b}) display a similar trend for the curves under analysis in comparison to those from Fig.\ref{fig:Random_structural_polarization}\textbf{a} and \ref{fig:Random_structural_polarization}\textbf{b}, we see now that consensus is only possible after a significant decay in the initial modularity of the network, see Figure \ref{fig:Ordered_structural_polarization}\textbf{c}, a task in which heterophilic rewiring preferences are more efficient than homophilic ones.

These results show that the rewiring process alone does not guarantee that the network population reaches consensus and, to achieve it, the rewiring process must occur with a sufficiently high frequency and be of the adequate type---heterophilic or homophilic--- to lead to the desired outcome. Further, the time to reach it differs even when the outcome is the same for both types. 

\begin{figure*}[!t]
    \includegraphics[width=\textwidth]{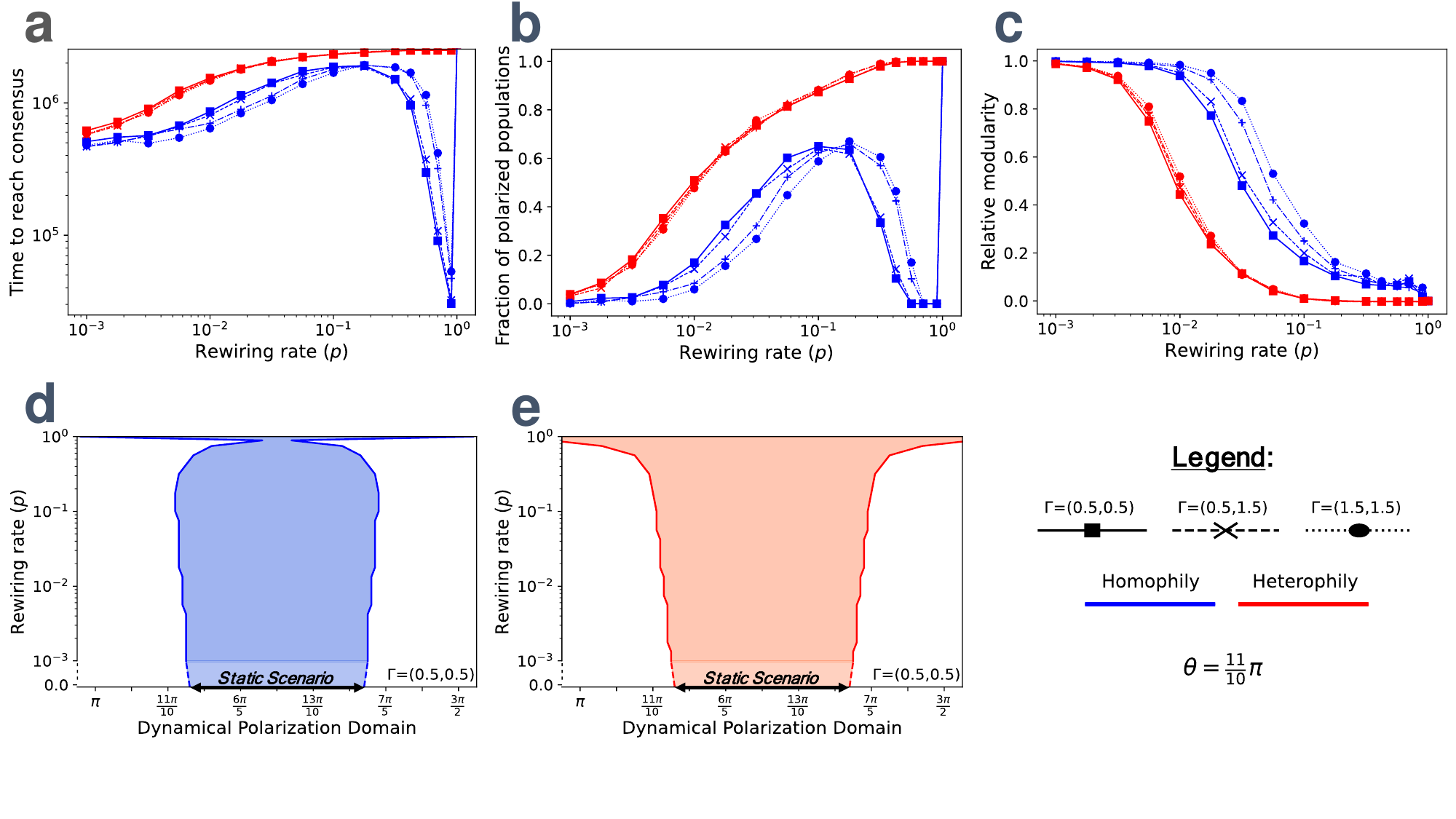} 
    \caption{\textbf{Opinion dynamics with Homophilic and Heterophilic rewiring preferences in a Polarization regime ($\theta=11\pi/10$)}. Panel \textbf{a} shows the fixation time as a function of the rewiring rate, $p$. Panel \textbf{b} shows the fraction of times a population ends in an opinion polarization state as a function of the rewiring rate, $p$. Panel \textbf{c} shows how the modularity of the initial structure decays  as a function of $p$. Panels \textbf{d})  and \textbf{e}), show the parameter range in which the observed dynamical pattern is aligned with the dynamical polarization of well-mixed populations. In panels \textbf{a}), \textbf{b}), and \textbf{c}) symbols indicate different values of the strictness of rewiring decisions ($\beta$) and colors the different rewiring preferences (homophilic in blue and heterophilic populations in red). Results are the average over $10^3$ independent simulations, each with an upper bound of $2.5 \times10^6$ iterations, on modular networks with $N=10^3$ nodes, an average degree of $\langle z_i \rangle=4 $.}
    \label{fig:Random_dynamical_polarization}
\end{figure*}


\section{Dynamic Polarization Regime}
Let us turn our attention to the regime when individuals easily change to rare strategies and associated with dynamic polarization in the static and well-mixed scenario. Non-complete social networks restrict the range of parameters in which dynamical polarization occurs, facilitating the formation of consensus. Although, in this case, the initial polarized structure of the population plays a less relevant role, it is important to understand to which extent rewiring can affect the chance and time to consensus.

Similarly to structural polarization, Figures \ref{fig:Random_structural_polarization}\textbf{a-c} suggest that the nature of the assessment (colors) seems to have a much more significant impact on the results obtained than the strictness (symbols) of the assessment itself. However, it is also possible to see that homophilic rewiring preferences are more sensitive to the strictness of the assessment, especially for intermediate values of $p$.

It is possible to recognize that the larger the rewiring rate, the easier it becomes for heterophilic populations to remain polarized, as evidenced in the increasing value of both the fraction of final polarized networks, Fig. \ref{fig:Random_dynamical_polarization}\textbf{b}, and the average time to reach consensus, Fig. \ref{fig:Random_dynamical_polarization}\textbf{a}. Most importantly, however, the rewiring probability can deeply change the dynamical pattern obtained by the final networks populated by homophilic individuals.

The time to reach consensus and the fraction of polarized populations, Figures \ref{fig:Random_dynamical_polarization}\textbf{a} and \ref{fig:Random_dynamical_polarization}\textbf{b}, tend to increase with $p$. This increase can be attributed to the lower probability of losing active links (i.e., links that can promote a change in opinion), resulting in a delay in fixation time. However, in populations with homophilic rewiring preferences, unlike heterophilic, fixation times start decreasing for large values of $p$, followed by a decrease in the fraction of populations that end in polarization. This unexpected non-linear behavior for homophilic rewiring preferences is fostered by the fact that if rewiring rates of homophilic individuals pass a critical point, rewiring will outpace opinion dynamics and, as such, foster the emergence of compact clusters of like-minded individuals in which the lack of variability of opinions in a neighborhood limits the changes of opinions spreading or, in other words, opinion updates. The same is not observed in heterophilic rewiring preferences, where individuals rewire their links, constantly looking to surround themselves with others of different opinions.

Moreover, the impact of rewiring dynamics on the initially modular structure of the social network is also worth analyzing. Like in the first case, the relative modularity decreases with increased rewiring probability ( Figure \ref{fig:Random_dynamical_polarization}\textbf{c}), but it exhibits a clearer S-shaped behavior and decays faster with $p$. This non-linear relationship suggests that the social network, depending on the rewiring rate ($p$), will be either strongly modular (slow network adaptation) or lack modularity network (intermediate in heterophilic rewiring and fast in homophilic rewiring).

Finally, in regards to the results obtained for well-mixed populations, we see that, when rewiring dynamics is considered, each of the studied rewiring preferences leads to different dynamical responses: homophilic rewiring preferences maintain or decrease the range of complexity parameters in which dynamical polarization is observed (Figure \ref{fig:Random_dynamical_polarization}\textbf{d}), but heterophilic rewiring preferences expand the range of parameters and in the limit of very fast rewiring rates match the well-mixed scenario.

\section{Conclusions}
This study delves into the dynamics of consensus formation in polarized social networks through a coevolutionary model that integrates competing opinions with adaptive network dynamics. Our research focuses on the interplay between homophilic and heterophilic rewiring preferences across a range of competing processes to enhance the understanding of social mechanisms that either facilitate or impede recovery from social polarization. In scenarios where information needs significant reinforcement to outcompete alternatives, our findings reveal two distinct pathways to consensus: a rapid one through independent community consensus and a slower one driven by rewiring dynamics. We demonstrate that heterophilic preferences are more effective in bridging communities for consensus in complex-information-diffusion contexts due to their ability to diversify opinion spaces. Conversely, in contagions requiring minimal reinforcement, homophilic rewiring emerges as more adept at fostering consensus and mitigating polarization risks, displaying a non-linear relationship with the rewiring rate. This finding suggests that homophilic preferences may be more resilient to polarization in environments where information bits are easily interchangeable.

Our research extends beyond the existing literature by examining how different rewiring preferences can mitigate or amplify polarization. This approach contrasts with studies that predominantly emphasize homophilic tendencies in social networks, suggesting a more nuanced role for heterophilic interactions in bridging divided communities.

The implications of our findings are significant for policymakers and designers of social or organizational networks, online and offline. By promoting a diversity of rewiring preferences, societies can enhance their resilience to the impacts of social polarization across a spectrum of simple to complex contagion processes. Identifying the specific complexity of the contagion of critical issues could further refine strategic approaches.

Expanding the proposed coevolutionary model to include more than two competing opinions and network communities is a natural next step as the dynamics become increasingly complex in such scenarios. Investigating different rewiring mechanisms, formulating optimal strategies for opinion dissemination, and designing targeted social interventions to control the spread of particular viewpoints are crucial areas for future exploration. Motivated by specific datasets, these extensions will allow us to fully capture the complexities of specific real-world social interactions and individual decision-making processes. Furthermore, developing new metrics to quantify social polarization among competing contagion processes, rather than standalone issues, will enable a deeper understanding of these complex polarization patterns. This expansion of research will provide a more comprehensive understanding of social dynamics and guide the development of effective strategies to address the challenges posed by social polarization.

\subsection{Acknowledgments}
\begin{acknowledgments}
FLP acknowledges the financial support provided by FCT Portugal under the project UIDB/04152/2020 -- Centro de Investigação em Gestão de Informação (MagIC). VVV acknowledges funding from ENLENS under the project "The Cost of Large-Scale Transitions: Introducing Effective Targeted Incentives."
\end{acknowledgments}

\bibliographystyle{unsrt}
\bibliography{main}

\end{document}